\documentclass{sig-alternate}

\begin{document}
%

\title{When Augmented Reality Meets Big Data}
%
%
%
%
%


\author{Zhanpeng Huang$^{\dag}$ \quad Pan Hui$^{\dag}$ \quad Christoph Peylo$^{\ddag}$\\
$^{\dag}$ HKUST-DT System and Media Laboratory,\\
Hong Kong University of Science and Technology, Hong Kong\\
\{soaroc, panhui\}@ust.hk\\
$^{\ddag}$ Telekom Innovation Laboratories, Berlin, Germany\\
christoph.peylo@telelcom.de}

\maketitle
\begin{abstract}
With computing and sensing woven into the fabric of everyday life, we live in an era where we are awash in a flood of data from which we can gain rich insights. Augmented reality (AR) is able to collect and help analyze the growing torrent of data about user engagement metrics within our personal mobile and wearable devices. This enables us to blend information from our senses and the digitalized world in a myriad of ways that was not possible before. AR and big data have a logical maturity that inevitably converge them. The tread of harnessing AR and big data to breed new interesting applications is starting to have a tangible presence. In this paper, we explore the potential to capture value from the marriage between AR and big data technologies, following with several challenges that must be addressed to fully realize this potential.
\end{abstract}




\section{Introduction}
Human intuition tells us it will be easier to make sense of and interact with information if it is merged with the physical world \cite{MacIntyre13}. As a modality to display information by overlaying virtual content on the current view of the world around us, AR enhances the way we acquire, understand, and display information without distraction from the real world \cite{Huang13}. Over the past few years, AR has been progressing by leaps and bounds in terms of technology and its applications. \par

However, even though technologies such as sensing, tracking, and displaying improve, AR application patterns broadly follow the lines of prototypes and demonstrations, such as virtual pop-up objects on 2D markers, what is inside the box, and sample data visualization \cite{Mathai14}. A major impediment to AR adoption is the lack of data sources described by MacIntyre et al. \cite{MacIntyre13} as ``walled gardens''. AR is data-hungry and requires more data than most applications. Apart of the application-specific content that users see and interact with, AR needs to feed applications knowledge about user surroundings and  descriptions about how it relates to the application data. Imperative environmental information may include geospatial coordinates and models of nearby buildings, features and semantical descriptions of objects, and linkage between real and virtual content. Previous works acquire data from the legacy database, but the data may be incomplete or out-of-date due to sparse sensing and the absence of persistent maintenance.\par

\fbox{\begin{minipage}{23.5em}
The first AR prototype was developed by Sutherland in 1960s \cite{Sutherland68}, but it is only in the last two decades that it has attracted a variety of attentions. A typical AR includes three characteristics defined by Azuma \cite{Azuma97}:
\begin{itemize}
  \item Combines the real and the virtual
  \item Interactive in real time
  \item Registered in 3-D
\end{itemize}
A way to supplement rather than replace real world as virtual reality (VR) with virtual content makes it preferable for applications such as tourism \cite{Schmalstieg07}, advertisement \cite{Insider12}, education \cite{Freitas08}, and assembly \cite{Henderson11}. Recently AR has grown popularity on mobile devices as a mobile AR (MAR). According to Juniper Research reports \cite{juniper09}\cite{juniper12}, the mobile advertising market for MAR-based apps has been estimated up to \$7.32 million by 2014 and MAR mobile applications will lead to almost 2.5 billion annual downloads worldwide by 2017.
\end{minipage}}\par

The high penetration of technologies spanning mobility, social networks, and the Internet of Things (IoTs) has catapulted the world in the era of big data. Mobile devices, social networks, online transactions, and instrumented machinery produce zettabytes of data as a by-product of their original operations, which can be analyzed to gain rich values not available from small datasets. Touted as a game changer, big data has been identified by the US government as a research frontier that is accelerating progress across a wide range of priorities \cite{PCAST10}.\par

AR and big data have been around and shaping their own landscapes in various fields for a few years, however, the intersection of two disruptive technologies has not attracted much attention yet. The rich insight of big data and novel display modality of AR is promoting the convergence of AR and big data. AR has great opportunities to bring innovation to big data in terms of visualization and interaction. Big data provides rich information for AR to breed new applications such as personal recommendations and intelligent information assistance. As AR applications normally generate massive data, which can be analyzed to further improve application performance and user experience, in this paper, we explore the potential opportunities from convergence of the two disruptive technologies alongside several challenging problems that should be addressed.

\fbox{\begin{minipage}{23.5em}
Big data is generally used to describe the exponential growth of data that is too large and complex to process using existing tools. The common definition of big data is articulated by Doug Laney as 3Vs \cite{Laney01}:
\begin{itemize}
  \item Volume: Data volume is too large to deal with traditional technologies.
  \item Velocity: Data is streaming in and out at high speed and must be processed within a timely way.
  \item Variety: Data from a wide range of sources are structured, semi-structured, and unstructured.
\end{itemize}
About 2.5 exabytes ($2.5\times10^{18}$) of data is generated every day \cite{IBM12}. According to a McKinsey report \cite{McKinsey11}, big data has the potential to reduce product development and assembly costs by 50\% and increases retailer's operating margins by 60\%. It is estimated to create a savings of 300 billion dollars in healthcare every year in the US alone. Big data analysis has been widely used in healthcare \cite{Consortium11a}, energy saving \cite{Council11}, financial risk analysis \cite{Flood11}, and so on.
\end{minipage}}\par

\section{AR-powered Big Data}
Interpretation is a major phase of the big data analysis pipeline, which is critical for users to extract actionable knowledge from massive and highly complex datasets. As an intuitional interpretation, visualization transforms plain and boring numbers into compelling stories to help users understand the data. In addition, big data analysis requires a human-in-the-loop collaboration as input at all stages of the analysis pipeline \cite{Labrinidis12}. Powerful interaction helps to explore and understand the data more easily and fully.

\subsection{Data Visualization}
With visualization, we turn the big data into a landscape that we can explore with our eyes. A visual information map is useful when we are drowning in information. Data visualization has the ability to take the complex as abstract symbols and increase the dimensions so that we can quickly understand. Visualization-based data discovery tools have already delivered greater customer and market insights to businesses around the world \cite{Intel13}. It is estimated by Gartner that data visualization tools will promote a 30\% compound annual growth rate in 2015 \cite{Sommer11}. \par

\begin{figure}[h]
\centerline{\includegraphics[width=8.0cm]{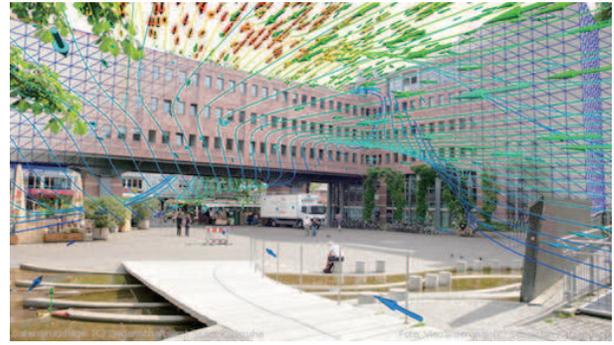}}
\caption{Visualization of a numerical flow field with real buildings makes the influence of the building on wind movement easily understood. (source: http://emcl.iwr.uni-heidelberg.de/research\_sv.html)}
\label{fig:1}
\end{figure}

\begin{figure}[h]
\centerline{\includegraphics[width=8.0cm]{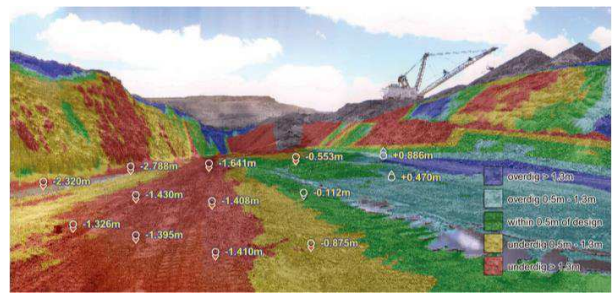}}
\caption{Excavation progress is overlaid on real scene to be compared against designs. (source: http://www.maptek.com/)}
\label{fig:2}
\end{figure}

In the past, we have employed massive forms for data visualization using tabular presentations, interactive bubble charts, treemaps, heatmaps, 3D data landscapes, and other types of graphics \cite{MacIntyre14a}. The data is displayed on flat mediums such as desktops and more recent mobile screens, which separates the visualization from the data source and user context. For instance, a virtual array of gauges to display temperatures would fail to fully describe the spatial distribution as the temperatures are accurate in relation to real world objects in a physical user context. In many cases, only if the data visualization is embedded in the physical world it is easier to gain more insights from the data.\par

Big data visualization can be enhanced if an AR layer is overlaid on real-time streaming data or user context. The AR enables users to be immersed in a world mixed with interactive data tags and blobs around their views. The data will be much more easily understood if it corresponds explicitly to real content. Since data is directly connected with physical context, display simplification which impairs user's understanding in traditional display is no longer required. For instance, to find a book from millions of collections in a library, traditional methods require digital 2D or 3D maps and floating bubbles for coarse positioning. Users may get confused as the digital map does not relate to users' current views. Powered with AR, users are able to ``see through'' walls and shelves to look for indications, e.g. the highlighting contour of the book. As an indication is usually located corresponding to a user's current view, it will be much easier to find the book. Collaborating with AR, big data is especially suitable for in-situ visualization and field diagnosis. For example, with rapid growth in information-rich building information modeling (BIM), AR has reached the construction and maintenance throughout the life cycle of a building \cite{Wilcox12}. This enables field workers to view the on-site feasibility design and assist construction with virtual simulations and clash detection. It also facilitates assets management in daily inspection activities based on a torrent of data from in-built sensors\cite{Irizarry12}. \par

To achieve success with big data visualization, a fresh rethinking about how to mix digital data with physical world and presenting the information to users is crucial. Apart from application-specific requirements, other considerations should be kept in mind. Floating bubbles are widely used by many AR applications, however, it seems to be pointless and no improvement on a 2D map \cite{MacIntyre14a}, especially when the data content can not be seamlessly integrated into the real world. Content should be merged into the physical world in a way that users perceive it as a real counterpart, just as the ``1st and Ten'' system used by ESPN in American professional football broadcast. To achieve this, visualization requires immense graphic effort and additional information for visual occlusion (e.g. something hidden behind a physical building) and ambient lighting.

\subsection{User Interaction}
Today's world is becoming a canvas for big data from a wide range of sources, which profoundly influences people's perception and understanding of the environment around them. It requires a user-friendly interface to interact with the digital world. Traditional user interface design is constrained by finite physical dimensions. It becomes especially severe when mobile devices are preferred as a medium to interact with torrents of data from social media, online transactions, and telecommunications. Simply enlarging the physical dimensions is not feasible as it increases device intrusion which results in a frustrating user experience.\par

As a new interface that straddles the digital and the real, AR mixes physical and digital content to create an imaginary, three-dimensional user interface without dragging you away from reality. The intangible interface with unlimited virtual space provides multiple ways for users to interact with more information. The ability to associate data with the physical world discloses the causality between data and reality. In addition, mashing up data from various sources dramatically increases the probability of discovering relevant and interesting things. A few works \cite{Heun13} further employed AR as user interface to redefine and attach new functions to physical objects. It is even more compelling when AR is integrated with accessories such as AR glasses and AR contact lenses. The wearable and invisible accessories reduce device intrusion, which provides a hand-free interaction experience. Google Glass has made huge strides in this area. Working with implanted and invisible accessories, AR allows users to interact with data securely. As the user interface is only seen and manipulated by the user alone, it reduces the risk of data and privacy leakage.\par

Data can be viewed through the AR interface as personal media or by multiple users in a collaborative way. As a personal information center, it collects data from various sources, and displays it on intangible interface without physical constraints. In the collaborative mode,  multiple users share the same data set and view it from their own angle. Each user can also probe into subsets respectively without interference.

Due to the high sensitivity of human eyes and the extremely high-speed conduit between the eyes and the brain, it is promising to use the eyes directly as an AR-based user interface \cite{Chapman10}. The University of Washington has developed a prototype of contact lenses crafted with built-in LEDs for AR display \cite{Parviz08}. Currently most AR exploration focuses the visual, but with the rapid advances in haptic, gustatory, and olfactory technologies, future AR interface will eventually be able to provide ``multimodal, multimedia experiences'' \cite{Kipper12}.

\begin{figure}[h]
\centerline{\includegraphics[width=8.5cm]{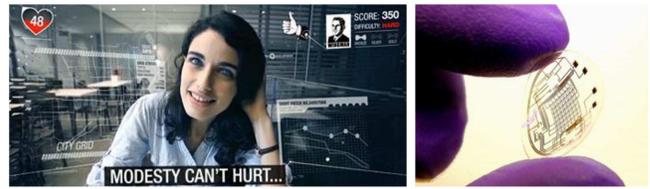}}
\caption{Left: ``Sight'', a short futuristic film by Eran May-raz and Daniel Lazo to look into a future AR with retinal lenses. Data from sensors, apps, and Internet augment current views (source: http://vimeo.com/46304267); Right: A prototype of bionic contact lenses with a built-in LED for virtual display \cite{Parviz08}.}
\label{fig:3}
\end{figure}

\begin{figure}[h]
\centerline{\includegraphics[width=8.0cm]{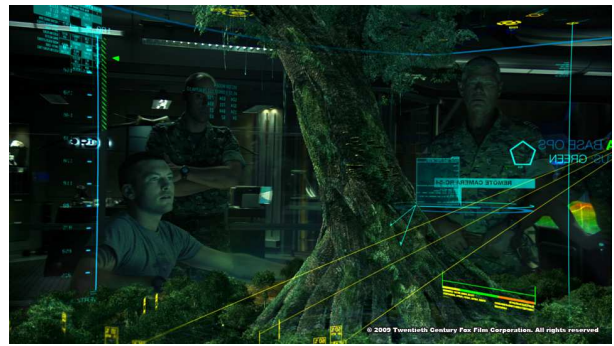}}
\caption{Large data visualization and interaction among multiple users in the science fiction movie ``Avatar'' directed by James Cameron, which may be portrayal of future user interface with AR. (source: http://www.avatarmovie.com/index.html)}
\label{fig:4}
\end{figure}

\section{Big-data-driven AR}
\begin{figure}[h]
\centerline{\includegraphics[width=8.5cm]{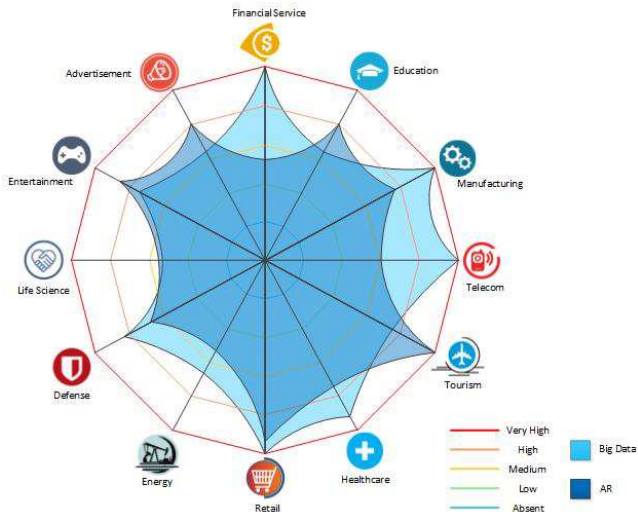}}
\caption{The influence circles of big data and AR on various fields. The influence on each field is qualitatively classified into five levels: very high, high, medium, low, and absent.}
\label{fig:5}
\end{figure}

Big data and AR have shaped new business regimes that were irrelevant in the past, but the landscape is undergoing a seismic shift with advances in technology convergence and connectivity. A majority of AR applications are constrained in a closed or prepared environment due to a limited dataset that is either not available or too sparse to use. As rapid penetration of mobile devices, social networks, and IoTs are generating considerable amounts of data, it makes sense that big data will enable AR to be more feasible for practical use. Figure \ref{fig:5} illustrates the influence of big data and AR to different fields. We select several of the most promising services that will be greatly promoted by big-data-powered AR in the near future.\par

\subsection{Retail}
AR has been in use in the retail business for a few years now. A majority of applications such as Junaio and Wikitude AR browsers overlay geospatial-related data on current view to offer general information. A few AR apps promote virtual advertisements to catch customers' eyes. However, it is frustrating to boost customers' interest in products if their behaviors and preferences are not taken into account. Without adequate information from customers, AR is less attractive for practical use and more like a gaudy, flashy technology.\par

The mobility trend preferred for purchase and social communication has led to the birth of digital consumers \cite{Infosys11}, which also brings us to the era of product digitalization. Digitally active consumers have changed their mode of transaction, communication, and purchase decision making. Consumers use mobile devices for online transactions, and social media for advice on what to buy and where to shop, leaving trails of their performances, states, and decision. These information are aggregated and complied to create a kind of digital self. It is obviously valuable to explore the large amounts of data to understand consumers' shopping preferences and behavior, which helps to tailor promotions and recommendations to customers. \par

However, it can be even more promising when it is combined with AR technology. Harnessed with big data, AR promotes vertical retail to individual consumers. Product information such as price and date can be overlaid on physical products for customers to get a quick overview of a product. It also supports customers in locating a product quickly using the ``X-Ray vision'' ability to look through other products to see a specific one behind. Backed by rich information from big data, AR displays the right product recommendation and personalized advertisement to augment customers' shopping context. A conscious and activated shopping context will impact on customers' mental presentation \cite{Dellaerta08}, which helps customers to be active, informed and assertive in their shopping decisions. \par

When the new eye gazing and facial expression technologies, and other physiological measurements eventually gain a tangible presence, it will enable us to better understand customers' focus and emotions to provide more accurate recommendations and advertisements. A few works have used eye-tracking glasses to collect customer point of gaze information for shopping behavior analysis. According to a Marks \& Spencer report \cite{Wood12}, people using mobile shopping channels spend eight times as much as people shopping in stores. Meeting consumers where they are is the key to future consumer engagement \cite{Udhas13}. With the advantages of high mobility and consistent virtual content provided by big data, AR breaks physical constrains to enhance the virtual shopping experience anywhere and anytime.

\begin{figure}[h]
\centerline{\includegraphics[width=8.5cm]{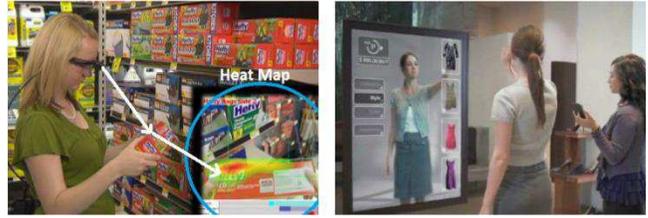}}
\caption{Left: a prototype to track a customer's point of gaze information in a shopping trip. (source: http://www.asleyetracking.com/); Right: Cisco's vision of  future shopping enhanced with AR. (source: https://www.youtube.com/watch?v=XM9ZOWPeiAk)}
\label{fig:6}
\end{figure}

\subsection{Tourism}
In terms of environmental awareness, AR is concerned about presenting contextual information and assisting in daily activities, which is particularly helpful when people are unfamiliar with the environment around them. By highlighting interesting features or bringing history to life, AR provides intuitive means to enhance a touring experience. As travel is normally associated with geo-spatial exploration, most AR applications for travel guides are based on geo-spatial information. A user's position is tracked using GPS and built-in sensors, which is then used to search and locate multimedia information from data sources such as point of interest (POI) databases, geocoded Tweets, and Flickr. Retrieved content including texts, images, and videos is rendered as floating bubbles and directly superposed on the current view. However, a cluster of bobbling tags, not aligned with anything, nor blending with the environment, nor hidden behind physical objects, seem not interesting, unhelpful, and not better than simply displaying the data on a 2D map \cite{MacIntyre14a}. AR applications will be just a curiosity until virtual content is seamlessly blended into the physical world, like virtual information about a cafe integrated into the facade as if it was attached to the cafe building.\par

Trends such as the fast deployment of sensor networks and the growing use of mobile devices and social networks are generating large amounts of both structured and unstructured data. The world is being digitalized as a complex digital web of locations, descriptions, photographs, and videos, and the status updates in a self-organizing manner with billions of individuals using mobile devices and logging onto various social networks. Aggregating and compiling the redundant fragmented data helps us to build a detailed and complete environmental model, which enables AR to understand users' surroundings better even in an open and unfamiliar environment. To give an example, Google Earth allows individuals to contribute digital 3D counterparts of real constructions to the community, which is building a 3D environmental model on a global scale in a crowdsourcing way. Using labels and geo-spatial coordinates, the geometry model can be fused with multimedia information and user-generated content (UGC) from shared platforms to enable virtual content generation and alignment with the physical world. Geometric models of historical constructions that have been demolished can also be created and then overlaid on the view of original sites with AR technology.\par

According to the Business Insider's survey \cite{Polland12}, intelligent recommendation is regarded as the most attractive expectation of tourism. As the world is getting ``smarter'' with big data, it comes the ability of tracking and measuring tourists' needs and behaviors to ensure a responsive and intelligent trip experience. For instance, personalized travel guide information is overlaid on the tourists' current view to avoid being distracted from tourist spots and getting lost. Native-language signs are automatically translated into readable words which are overlaid on the original places. To go a step further, information such as locations of nearby rest sites and restaurants can be recommended according to tourists' needs based on walking distance and time. The advantage of merging the virtual and the real also enables AR to turn the tour into a game. Google's pervasive game Ingress establishes virtual portals at interesting points such as public artworks, landmarks, and cenotaphs, which can be used to help tourists uncover facts about the tourist spots like a treasure hunt.

\begin{figure}[h]
\centerline{\includegraphics[width=8.0cm]{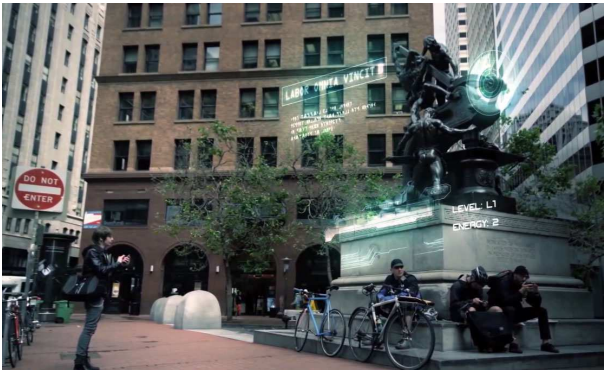}}
\caption{A trailer of the pervasive game Ingress. AR promotes gamification of travel to increases tourists' interest in tourist spots. (source: http://www.ingress.com/)}
\label{fig:7}
\end{figure}

\subsection{Health Care}
When we are making life and death decisions, immediate access to relevant and necessary information is of the utmost important. AR importance in healthcare industry is attributed to its ability to instantly in-situ display relevant information when required. AR has been used in the medical field for nearly ten years. We have seen AR's powerful ability of ``x-ray vision'', which has been used to provide contextual cues for diagnosing patients and learning tools for medical students by projecting CT scans or medical images on a current view. In one example, images of veins are overlaid on a nurse's view of a patient's hand to help the nurse insert the IV in one painless attempt. In another AR application developed by German research institute Fraunhofer \cite{Bimmer13}, a digital overlay of key blood vessels are displayed on the iPad when the doctor holds the embedded camera on a patient's body to avoid accidentally cutting them. Although the early examples have proven AR's ability to change the healthcare landscape, AR's great lifesaving potential for healthcare industry can not be fully presented without big data support. Without adequate data sets, AR is merely for medical education purposes and a bedside manner test. A decision is strongly made based on doctors' hunches and experience rather that the data itself.\par

According to a survey by Manhattan Research \cite{Research13}, 72\% of physicians routinely use computer tablets every day. With substantially increasing usage of tablets among physicians and digital care devices among patients, paper prescriptions and manual health records are being replaced by electronic health records (EHRs). Patient data is being digitalized, leading to a flood of digital data from which medical decisions that previously were based on guesswork and experience can be made based on data itself. Creating a virtual viewfinder to display a pertinent health record enables the doctor to quickly access valuable information in the context of patients. In-suit visualization of historical illnesses or tissue damage over patients themselves helps the doctor understand his or her patients better. AR can even enhance the decision making process by scanning for problems and providing an immediate field diagnosis when CT scans and images of the patient's symptoms are taken. Using AR to augment live-streamed video of remote patients with EHRs, the doctor is able to provide a private remote face-to-face diagnostic service to patients, which is especially helpful for rural regions without enough medical personnel. In future, AR can even visualize a virtual operating room for doctors at different places to diagnose a patient in a collaborative way. \par

Working closer with ourselves in daily life, AR can be significantly important for self-tracking health, as wearable devices that could sense heart rate, blood oxygen, or even cholesterol level become available to everybody. With each of us becoming a walking data generator \cite{McAfee14}, we would be able to keep track of our own basic health statistics. AR displays real-time notifications to allow us to immediately understand our own health condition, and it may even make suggestions based on health statistics and diet. Image in the near future, when you want to eat a hamburger, the nutritional information is displayed on a digital overlay around the hamburger, followed by a friendly reminder that your cholesterol level may not like it. \par

\begin{figure}[h]
\centerline{\includegraphics[width=8.0cm]{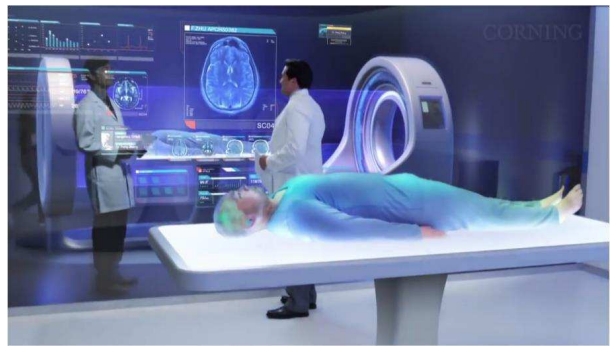}}
\caption{Corning's vision of a future operation room augmented with remote presence. (source: https://www.youtube.com/watch?v=VF2mKavngHE)}
\label{fig:8}
\end{figure}

\subsection{Public Services}
The government is responsible for providing public services, which both produce and consume a large amounts of data. As the largest spender in any economy, the government has the most diverse channels for collecting data from public services, which includes transportation, social services, national security, defense, and environmental stewardship, among others. According to a joint global survey by Bloomberg Businessweek and SAP \cite{Mullich13}, 81\% of the questioned believe that public sector will inevitably be transformed by big data. The government can use big data to provide better services to citizens. Public services, such as transportation and security protection, will be more efficient and productive if they are delivered directly to individuals in their own contexts with AR technology. For instance, the upcoming traffic flow displayed on a screen will help drivers avoid traffic accident. Personal information overlaid on passengers will enable security specialists to very quickly verify identification and reduce screening traffic. ``Augmented government'' \cite{Doolin13} has been proposed to improve government services with AR technology in a few US public sectors. The AR strategy for government service delivery will be more promising if it is supported by big data from surveillance systems, mobile and sensor networks, and social media.\par

As sensor and wireless networks continue covering vehicles, it is much easier to collect massive traffic stats to make us aware of the traffic situation. For example, cars present within a range of distance can share GPS positions, speed, and direction information with a vehicular ad hoc network (VANET). AR can display the information in front of drivers for performing thread assessment and predicting any potential car crashes. In particular, by harnessing the ``x-ray vision'' capability, drivers can ``see through'' buildings or vehicles to watch for vehicles positioned in their blind spots. Overlaying essential information such as driver's license and vehicle's location and speed directly over the vehicle will also help traffic police to quickly determine whether the driver violated any traffic rules. A similar strategy can be employed by security agencies to rapidly identify suspects.\par

The city is being digitalized by ubiquitous sensor and social networks, which makes it ``transparent'' for city managers to look into. To give an example, in a civil engineering maintenance work scenario, a virtual image of a subsurface infrastructure can be superimposed on a field workers' vision of the site to enable them rapid perception of the underground network layout. Field workers can collaborate from different aspects by giving contextualized views to each supporting role \cite{Mathai14}. Individual view is personalized and annotated for each worker's context, such as electrical-line view for the electrician and plumbing-line view for the plumber, which harnesses the collective intelligence of all roles to improve efficiency. In another example, a virtual bird's eye view directly overlaid on an emergency staff's vision will greatly assist in the search and rescue of persons trapped in a burning or collapsed building. As civil infrastructure such as electrical and water supply networks are getting ``smart'' with IoTs, torrents of data can be used for in-suit visual analysis without field damage by using AR technology. \par

\begin{figure}[h]
\centerline{\includegraphics[width=8.0cm]{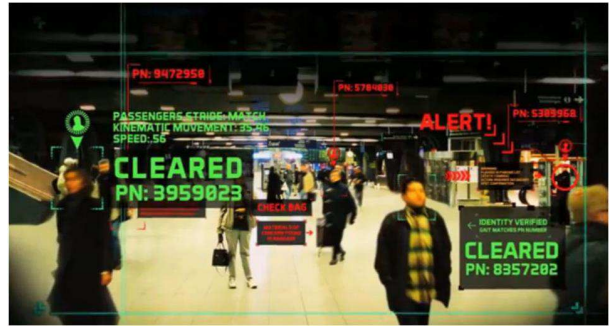}}
\caption{In future, an analyzed personal profile is overlaid on an agency's field of vision for fast security screening without direct contact \cite{Doolin13}.}
\label{fig:9}
\end{figure}

\section{Challenges}
We have seen an emerging shift in mindsets of big data and AR from industry and business, but there are still significant barriers to overcome. Several barriers, such as intrusive display, battery life, and highly fragmented data, are practical. Some barriers are conceptual. We should first address these challenges before we can see their full potential in action.\par

Converging big data and AR brings practical technical challenges from both sides. Heterogeneity, scale, and complexity are problems with big data that impede the process of all phases from data acquisition to aggregation and analysis. AR applications are also constrained to extensive calibration, incomplete reference modeling, and poor environmental sensing. A comprehensive discussion of the technical problems from each side is out of the scope of this paper. Interested readers can instead refer to \cite{Huang13} and \cite{Labrinidis12} for details. Herein we explore emerging practical technologic problems and common conceptual barriers induced by converging both technologies. \par

\subsection{Timeliness}
AR applications generally require real-time performance to guarantee fluent user interaction, which requires immediate analysis results. However, large-scale data analysis usually takes so long due to the voluminous and highly fragmented data. The problem becomes considerably more challenging when the trend of minimization in AR devices conflicts with the growing volume and fragmentation of big data. \par

Incrementally computing a small amount of new data based on partial results in advance can get a quick determination, while the crowding new data and new analysis criteria may render the results invalid. As cloud naturally fits for big data and AR, a dramatic shift has been moving towards cloud computing. With a theoretically infinite computing capability and memory capacity, cloud is able to store a large amount of data and handle computationally intensive tasks within a fixed time cap. A few applications \cite{Huang14} have proven cloud's ability to meet requirements in a timely fashion. In addition, offloading computation and data storage enables client-side AR devices to be small and sustainable enough without intrusion.

\subsection{Interpretation}
By integrating big data with AR, we acquire data analysis ability. However, the ability to analyze data is of limited value if the results can not be understood by AR. Users of big data analytical systems are data scientists while AR users are customers without much background knowledge. The present data analysis pipeline has rather been designed explicitly to have a human in the loop because many patterns are obvious for humans but difficult for computer to understand \cite{Labrinidis12}, however AR prefers to be intuitive and used without interruption. Analysis results require interpretation in the context of AR. For instance, the output of a customer behavior analysis system is normally customer stats, but AR is responsible for how to use the stats. AR should be able to interpret the results as preferential information so as to provide a recommendation to a customer's specific context. Although big data is good at discovering correlations, especially subtle corrections that are not possible from a small data set, it does not tell us which correlations are meaningful, while AR requires semantically meaningful information to relate to the users' context.\par

There is no easy way for big data and AR to intelligently interpret for each other, but a collaborative effort to embrace AR content and provide native APIs for AR to interpret semantically-tagged data from all data generators is a possible solution. A standard data format such as Augmented Reality Markeup Language (ARML) \cite{MacIntyre13} is an essential step in the right direction. \par

\subsection{Privacy}
There is a growing concern about privacy in the context of both AR and big data. In order to provide a personalized recommendation, AR is generally required to access users' personal information and record their location. It has been proven that users' identities and their movement patterns have a close correlation \cite{Gonzalez08}. Even though people pay attention to personal information, an attacker can infer private information from their location information \cite{Labrinidis12}. Hiding location is more challenging than hiding private information as people do not usually care about their locations leakage. The privacy concern is becoming more serious as maintaining a digital self on social networks becomes an active and voluntary process. People are encouraged to share their information seamlessly across social networks and mobile platforms. Although the data is fragmented and incomplete from individual sources, it is interrelated and includes frequent patterns and redundant knowledge. When it is aggregated from numerous data sources, hidden relationships and models can be extracted by data analysis. \par

Privacy is both a technological and sociological problem. Forceful laws and regulations may be required to avoid inappropriate use of personal data and malicious AR applications. From the technical standpoint, differential privacy is a possible way of accessing data with a limited privacy risk, however the information is reduced too far to be useful in practice. Not only that, it is ill-suited for dynamically changing data.


\section{Conclusion}

In this paper rather than just an up-to-date survey of big data and AR technologies, we have broadened our outlook to merge them to breed new applications. The factors for promoting the convergence of the two technologies also created some of challenges not previously experienced by either technology. The key principle is to combine healthy features of each technology as we devise novel applications using both. Although the technologies are still in their infancy, attention has always been a necessary component of the convergent strategy of AR and big data. While the concepts are not unfamiliar to us, most participants have not yet harnessed their full potential. Only when we take full action can we get a real picture of what it will take for big data and AR to move from being just hype to becoming real game changers.

%
\bibliographystyle{abbrv}
\bibliography{sigproc}  
%
%
\end{document}